# Global organization of protein complexome in the yeast *Saccharomyces cerevisiae*

**Sang Hoon Lee[1], Pan-Jun Kim[2], Hawoong Jeong[3*]**

[1] IceLab, Department of Physics, Umeå University, 901 87 Umeå, Sweden

[2] Institute for Genomic Biology, University of Illinois at Urbana-Champaign, Urbana, IL 61801, USA

[3] Institute for the BioCentury and Department of Physics, Korea Advanced Institute of Science and Technology, Daejeon 305-701, Korea

*Corresponding author

Email addresses:

    SHL: sanghoon.lee@physics.umu.se

    PJK: extutor@illinois.edu

    HJ: hjeong@kaist.edu

# Abstract

**Background**

Proteins in organisms, rather than act alone, usually form protein complexes to perform cellular functions. We analyze the topological network structure of protein complexes and their component proteins in the budding yeast in terms of the bipartite network and its projections, where the complexes and proteins are its two distinct components. Compared to conventional protein-protein interaction networks, the networks from the protein complexes show more homogeneous structures than those of the binary protein interactions, implying the formation of complexes that cause a relatively more uniform number of interaction partners. In addition, we suggest a new optimization method to determine the abundance and function of protein complexes, based on the information of their global organization. Estimating abundance and biological functions is of great importance for many researches, by providing a quantitative description of cell behaviors, instead of just a "catalogues" of the lists of protein interactions.

**Results**

With our new optimization method, we present genome-wide assignments of abundance and biological functions for complexes, as well as previously unknown abundance and functions of proteins, which can provide significant information for further investigations in proteomics. It is strongly supported by a number of biologically relevant examples, such as the relationship between the cytoskeleton proteins and signal transduction and the metabolic enzyme Eno2's involvement in the cell division process.

**Conclusions**

We believe that our methods and findings are applicable not only to the specific area of proteomics, but also to much broader areas of systems biology with the concept of optimization principle.

# Background

High-throughput detection of protein-protein interactions (PPIs) has long been one of the most intensively studied areas in systems biology. Currently, recent progress in experimental techniques, such as the yeast two-hybrid (Y2H) method or tandem-affinity-purification method coupled to mass spectrometry (TAP-MS) [1,2,3], has allowed the compilation of massive data set for protein interactions of the budding yeast, *Saccharomyces cerevisiae* (*S. cerevisiae*). From the vast amount of PPI data, protein interaction networks (PINs) have been vigorously investigated. In PIN, proteins are represented as nodes, and two proteins are linked if they interact with each other. Basic topological measures, such as a degree (the number of neighbors a node has), and their correlations are adopted to explain various properties of proteomes. For instance, a degree in PIN is the number of partner proteins with which an individual protein interacts. One of the earliest achievements is the heavy-tailed degree distribution (denoted as "scale-free" network) and its importance in characterizing the essentiality of proteins [4]. In addition, there have been further explorations on the negative degree-degree correlation profile ("disassortative" nature) related to the modular structure of protein interactome [5,6]. Models for PIN have been developed by incorporating such characteristics [7,8]. In this way, it has been witnessed during the last decade that studies on complex networks [9-12] play a great role in characterizing such interacting entities, and the term "network biology" was coined to refer to networks whose nodes correspond to proteins, metabolites, genes, etc [13].

In spite of such recent progress, there are shortcomings of that type of simplified network representation of PPI in that detailed mechanisms or higher structures of interactions are often neglected. Most of all, only the pairwise binary interactions between two proteins are considered in PPI, and they are not sufficient to capture real biological processes involving a stable form of proteins grouped according to their biological functions. For instance, two proteins A and B can be involved in a certain biological process by forming a linear chain with another protein C, such as A-C-B, without any direct contact between A and B. To overcome such problems, it is crucial to consider protein complexes, which are groups of proteins performing specific cellular tasks [14-18]. In this paper, we focus on protein complexes and their component proteins in *S. cerevisiae* whose data set has been established through the recently developed TAP-MS technique [3]. We apply the concept of bipartite network [9-12] to the protein complexes and proteins associated with them. A bipartite network is composed of two types of nodes, and there exist links only between nodes of different types, as usually found in social "affiliation" networks, where individuals are joined by common membership of groups [9]. In our case, protein complexes and their component proteins are the two types of nodes, and a complex is connected to a protein if the protein is one of the components of the complex, as shown in Fig. 1. We construct the bipartite network and obtain its one-mode projections [9], *i.e.*, the weighted network of individual proteins (protein-protein network) and the weighted network of complexes (complex-complex network), where the weight on each link is given by the number of shared complexes or proteins, respectively.

The weighted network is an extension of complex networks, with additional information, weight, on each link, and has been investigated in the context of network theory [19], including the metabolic network analysis [20]. From this weighted network analysis, we can extract quantitative topological characteristics of the interrelationship among protein complexes and component proteins. In this work, we analyze the distributions of degree and strength (an "extension" of degree, which is defined as the sum of weights on the links connected to each node). For both bipartite and projection networks, the distributions approximately follow the exponential distribution, while most PINs in the literature are claimed to show more inhomogeneous degree distributions [4,5,13,21]. Furthermore, two classes of proteins denoted as "core" proteins and "attachment" proteins are shown to play different roles in the complex formation.

Besides the aforementioned preliminary analysis on the structural properties of networks, as a main topic of our work, we suggest a new method for the systematic estimation on cellular abundance of protein complexes and the assignment of biological functions to them, as well as those of individual proteins. For further researches, such as kinetic modeling of the cell, estimating the abundances and functions of complexes and component proteins can provide a much more quantitative description of behaviors in a cell than a list of protein interactions. However, most of previous attempts about the assignment of unknown functions to proteins or complexes usually rely on *local* information. For example, there is a method mapping the problem to the local-similarity-based Potts model [22], which is an intrinsically stochastic method and depends on a temperature-like parameter. Another previous work is a statistical argument based on hypergeometric tests, as in Ref. [23], but it also requires a specific external threshold and additional corrections to fit the data to biological reality. In contrast, our deterministic method is based on the optimization problem related to the *global* organization of protein complexes, and yields a number of experimentally verifiable results. From the results of the optimization, we cross-analyze the condition-dependent abundance and functions of complexes, which also supports the reliability of our method as well as our extensive statistical validation process. All of these examples clearly show the effectiveness of our method, and we believe that this method has the potential to significantly stimulate further experimental studies.

## Methods

**Bipartite Network Representation of Protein Complexes**
Figure 1 is a schematic diagram of our bipartite network representation of protein complexes and their component proteins. The bipartite network shown in Fig. 1(a) consists of two types of nodes, *i.e.*, complex and protein, where a complex is linked to a protein if the protein is a component of the complex. In the protein-mode projection or protein-protein network [Fig. 1(b)] whose nodes are individual proteins, two proteins are connected if both of them are used to form at least one complex. The number of complexes they *share* in this manner is assigned as the weight of each link. The complex-mode projection (complex-complex network) described in Fig. 1(c) is obtained similarly, where complexes are linked if they include at least one protein in common, and the number of such shared proteins is the weight.

Note that the bipartite network itself is not a weighted network, while the protein-protein and complex-complex networks are weighted ones, based on the information provided by the bipartite network. Similar approaches have been presented in previous works [24,25], with older datasets [14,15]. For instance, Mashaghi *et al*. [24] used a similar approach, but we consider all three kinds of networks, while Ref. [24] dealt only with the bipartite network and the complex-mode projection. In addition, we use a weighted version of one-mode projections in this work, while the unweighted one-mode projection networks were considered in the previous work [25].

We analyze the basic statistical characteristics of these three networks, using the methodology derived from a decade of complex network studies [9-12,19]. The dataset of protein complexes and their component proteins in *S. cerevisiae* is listed in Ref. [16], where significant overlaps (shared component proteins) among protein complexes make it possible for us to analyze the interconnected network structure, and determine the abundance and functions. In Ref. [16], Gavin *et al*. use the TAP-MS technique [3,14] to generate the list of protein complexes and their component proteins. The proteins in each complex are classified as cores and attachments, according to their significance in the formation of specific complexes. We will discuss the different roles of core proteins and attachment proteins later on. The datasets consist of 491 protein complexes and 1,491 component proteins, among which 1,147 proteins participate in complexes as cores and 1,134 proteins as attachments. Therefore, 790 proteins are used as both cores and attachments.

**Estimation of Complexes' Abundance Based on Optimization**

Proteins interact with each other and form complexes in order to perform specific biological functions. Naturally each protein complex executes specific biological functions with its characteristic composition of component proteins, and the identification of such functions and abundance *in terms of complexes* is important. Unfortunately, such exact identification of complexes' functions and estimation of their abundance is far from being complete. Only heuristic methods, using local (binary) interaction of proteins, were used in previous studies [16,22,23], and there were some limitations, such as the arbitrarily specific values of threshold. In this section, we present our new optimization method to determine the abundance and function of protein complexes, based on the information of their *global* organization. Estimation of abundance would be especially crucial, and is directly applicable to various works, such as setting up kinetic models that involves proteins, by providing reaction coefficients.

Suppose that the copy number of protein $i$ ($i = 1,...,N$; $N$ is the number of proteins) and the number of complex $j$ ($j = 1,...,M$; $M$ is the number of complexes) are given by $p_i$ and $c_j$, respectively. Also, let us denote the number of protein $i$ in the complex $j$ as $S_{ij}$, where $S_{ij} = 0$ if the complex $j$ does not harbor the protein $i$ as its component. Then, in an ideal situation where *all the abundances or concentrations of proteins in a cell correspond to the exact amount to be used in forming a complex*, similar to the approach used for kinetic modeling in [26], the variable sets $\{p_i\}$ and $\{c_j\}$ satisfy

$$p_i = \sum_{j=1}^{M} S_{ij} c_j. \quad (1)$$

The question is how to determine $\{c_j\}$ (variables) with known values of $\{p_i\}$ and $\{S_{ij}\}$ (constants). However, since the number of proteins $N$ is larger than the number of complexes $M$, the set of linear equations above is *over-determined*, and we assume that in reality it is not feasible to satisfy all the equations in Eq. (1). In practice, therefore, we assume that the number of proteins in a cell should be greater than or equal to that necessary to form complexes, *i.e.*, $p_i \geq \sum_{j=1}^{M} S_{ij} c_j$, which is the basic constraint of our optimization scheme. Therefore, instead of looking for an exact solution satisfying Eq. (1), we try to minimize the deviation from the ideal situation in Eq. (1), given by

$$DA = \sum_{i=1}^{N} \left[ 1 - \sum_{j=1}^{M} (S_{ij} c_j) / p_i \right] \qquad (2)$$

where the summation is only for indices $i$ where $p_i > 0$. Now, for the given values of $p_i$ and $\{S_{ij}\}$, our basic strategy is to determine $c_j$ values that minimize $DA$ in Eq. (2), and this problem is numerically solved by the linear programming (LP) technique. Moreover, after the determination of $c_j$ values, if some values of $p_i$ are unknown, we can *assign* those values of $p_i$ using Eq. (1) for the ideal situation. This optimization is based on an assumption that organisms have been evolved in a way that increases efficiency by reducing wasted resources. We use the abundance data of yeast proteins $\{p_i\}$, in Ref. [27], because the datasets for both rich (YEPD) and minimal (SD) media are available, and we exploit those conditional changes of abundance later on for our analysis. The values of $\{S_{ij}\}$ are given by the list of protein complexes used in this work.

**Inference of Complexes' Biological Functions Based on Optimization**

For a functional annotation of complexes, we consider the following optimization scheme. Suppose $Fp_{ik} = 1$ if protein $i$ performs a function $k$ ($k = 1,...,F$; $F$ is the number of functions) and $Fp_{ik} = 0$ otherwise. Similarly, $Fc_{jk} = 1$ if complex $j$ performs a function $k$ and $Fc_{jk} = 0$ otherwise. The organization of protein complexes is given by $U_{ij}$, where $U_{ij} = 1$ if protein $i$ is a component of complex $j$ and $U_{ij} = 0$ otherwise (note that $U_{ij} = 1$ if $S_{ij} > 0$ and $U_{ij} = 0$ if $S_{ij} = 0$). In this case, the constraint is given by

$$Fp_{ik} \leq \sum_{j=1}^{M} U_{ij} Fc_{jk}, \qquad (3)$$

meaning that *every function assigned to a protein must be assigned to at least one of the complexes in which the protein participates*, which is reasonable, based on the assumption that each biological function is usually expressed cooperatively in the form of protein complexes. We are aware that our dataset of proteins and Munich Information Center for Protein Sequences (MIPS) database [28] are not exhaustive, so the outcome is not a perfect functional catalogue. However, since our method itself is stable and resilient, as we demonstrate later on the validation part, the result will become more accurate as the input dataset becomes more reliable. Even if there are some errors and exceptions, such as single proteins not being included in our complex datasets, we emphasize that, based on notable examples presented later, our method clearly produces biologically significant results.

Our criterion used for the optimization process is to assign the most *definite* functions to each complex. In other words, we try to find functions that are *inevitably* assigned among all the other solutions satisfying the constraint (3), in the "safest" way. In the spirit of parsimony, we minimize the number of complexes assigned with each function. Mathematically, this corresponds to minimizing the following quantity for each $k$:

$$DF_k = \sum_{i=1}^{N}\left[\sum_{j=1}^{M}\left(U_{ij}Fc_{jk}\right) - Fp_{ik}\right] . \qquad (4)$$

Therefore, our strategy of determining each protein complex's functions is determining $\{Fc_{jk}\}$ which minimizes either Eq. (4) under the constraint (3), with the given values of $\{Fp_{ik}\}$ and $\{U_{ij}\}$. Note that all variables and constants in this case are integer values, in contrast to the abundance estimation. This optimization problem is numerically solved with the mixed integer programming (MIP) technique. After determining the $\{Fc_{jk}\}$ set, similar to the case of abundance case, we can *conjecture* the function of protein $i$ as $k$, if $\sum_j U_{ij}Fc_{jk} \geq 1$ while $Fp_{ik} = 0$, because the protein $i$ is considered to perform the function $k$ by participating in at least one of the complexes to which the function is assigned.

The function assignment of proteins, *i.e.*, $\{Fp_{ik}\}$ is available from various sources, such as MIPS [28] or the Yeast Genome Database [29], and the values of $\{U_{ij}\}$ are given by the list of protein complexes used in this work so far [16]. We use the MIPS database for the initial function assignment of individual proteins, which is set $\{Fp_{ik}\}$, because its hierarchical organization of Functional Catalogue (FunCat) [28] helps the systematic interpretation of results in different hierarchical levels. We would like to emphasize that our method, for the first time, *systematically* assigns multiple numbers of functions to all the complexes in the genome-wide scale, without any free parameter or initial condition dependency, in contrast to other previous methods [22,23].

There is a certain degree of resemblance between our method and the message passing algorithm, such as the belief propagation (BP) [30,31], in inferring or assigning (previously unknown) properties to variables in bipartite graphs. For instance, the message from a variable node *v* to a factor node *u* in BP is analogous to assigning function or abundance of complexes (factor nodes in the complex-protein bipartite network), and the following process of message passing from *u* to *v* corresponds to assignment to proteins with previously unknown functions or abundance. We can even consider the configuration of object functions in Eqs. (2,4) as the "energy landscape" of constraint optimization problems, where methods like BP play important roles. In spite of this analogy, there are significant differences as well. First, our method is designed to deal with a specific type of problems composed of the set of factor nodes with no *a priori* information and the set of variable nodes with information except for a small fraction of unknown cases. Second, and more importantly, in contrast to BP, where the local flow of inference is used, our method is based on global optimization of object functions. In this respect, the previous works using local or Bayesian inference to assign protein functions [22,23] are similar to the message passing algorithm than our method. Of course, finding deeper mathematical analogy between our method and such traditional inference algorithms would be certainly interesting, but that would be beyond the scope of this paper.

## Results

**Statistical Properties of the Bipartite and One-Mode Projection Networks**
First, we construct the bipartite network without distinguishing between core and attachment proteins. As mentioned in the 'Methods' section, the bipartite network consists of 491 complexes and 1,491 proteins as its two different types of nodes. The average degree of complexes (the average number of proteins that a complex includes) is 13.41 (with standard deviation of 14.03), and the average degree of proteins (the average number of complexes in which a protein participates) is 4.42 (with standard deviation of 4.05). Figure 2 shows the degree distributions of the bipartite network, where we separate the distributions for complexes and proteins since they are different types of nodes. As shown from the fitted curves in Fig. 2, the degree distributions for both kinds of nodes are much closer to the exponential distribution $p(k) \sim \exp(-k/k_0)$ (except for the "tail" part from the finite-size effect) than the power-law distribution, because the cumulative distribution $P(k)$ of the exponential distribution $p(k) \sim \exp(-k/k_0)$ also follows the exponential distribution from the relation $P(k) = \int_{k}^{\infty} dk' p(k') \sim \exp(-k/k_0)$, in the continuum limit. Two complexes, denoted as Complexes 27 and 56 in Ref. [16], include prominently many component proteins, especially including proteins affiliated to the ribosome, as shown in Table 1. The two proteins, with the maximum connectivity 24 in the bipartite network, are Rps22a and Rpl36b, which are ribosomal proteins [29].

One-mode projections of our bipartite network represent interactions or relatedness among complexes and individual proteins. In the complex-mode projection, intuitively, two complexes are considered to be related if they share proteins, and the "intensity" of their relatedness, the number of proteins they share, is quantified as the weight. Similarly, in the protein-mode projection, two proteins are thought to be related if they participate in common complexes, due to the fact that proteins perform a biological function as a unit of the protein complex. Because more than two proteins can be cooperatively involved in the protein interactions, investigating protein interactions from the formation of complexes has its own advantages over other methods using only the pairwise interactions, such as the Y2H method [1,2]. The degree distributions of the complex-mode projection shown in Fig. 3(a) and the protein-mode projection in Fig. 3(b) show the exponential distribution. The strength $s_i$ of a node $i$ in a weighted network is defined as [19]

$$s_i = \sum_{j \in V_i} w_{ij}, \quad (5)$$

where $w_{ij}$ is the weight of the link connecting the node $i$ and $j$, and $V_i$ is the set of neighbor nodes of $i$. It can be considered as a natural extension of the degree of a node in weighted networks, and the strength distributions of our weighted complex-complex and protein-protein networks are shown in Fig. S1 (in Additional Files). The strength distributions are also closer to the exponential form, rather than the power-law distribution observed in many other biological and other real-world weighted networks [19,20]. It is true that the maximum degree (163) of complex-mode projection network is very large, considering the number of nodes (491), but this is merely due to the large connectivity (average degree) of distribution, not to the "scale-free" functional shape.

Although the notion of hub, referring to nodes with many connections to others, in the exponential degree distribution is not as strong as in the power-law degree distribution, we can still identify nodes with relatively large degrees and strengths by examining the "tail" part of the distributions. Table 1 lists the complexes and proteins with the ten largest degree and strength values in both the bipartite and projection networks. Again, the ribosome-associated complex (RAC) is a notable example annotated in the datasets [32]. Other unknown complexes with large degrees and strengths also include many ribosomal proteins (denoted as 'Rpl-') as their components. Some ribosomal proteins have the largest degrees and strengths in the protein-mode projection as well, e.g., Rpl33a, Rpl30, Rpl16b, and Rpl26b [29]. The fact that many complexes and proteins with largest degrees and strengths are affiliated to the ribosome can be interpreted in two ways. One hypothesis is that because the ribosome is involved in the production of proteins, ribosomal proteins might be promiscuously attached to many unspecific complexes. The other is that ribosomal complexes are usually large and composed of many ribosomal proteins, which implies the genuine property in this case. To determine which hypothesis is more plausible, we have manually checked the complexes in which those proteins are involved and found that most component proteins of those complexes are also ribosomal proteins, which supports the latter hypothesis. For instance, complexes 27, 55, 56, which have the three largest degree (number of component proteins) in the bipartite network, are all assigned with 'ribosomal proteins' from our optimization method. We note that in some previous works, the ribosomal proteins are considered as non-specific contaminants involved in promiscuous interactions and filtered out, for example, in Ref. [17]. However, our dataset from Ref. [16] already passed the stringent test based on socio-affinity index, taking into account the frequency of proteins within the dataset and naturally discriminating true from spurious interactions [16]. We believe that the criterion used in Ref. [16] can be more systematic and reasonable than the "3% rule" used in Ref. [17], so the statistics related to the ribosomal proteins may actually be reflective of the biological reality. However, it would also be meaningful to mention some non-ribosomal complex/proteins with such large degree/strength values. Therefore, we have identified other examples such as complexes assigned with tRNA modification function, fatty acid metabolism, and aromatic anabolism from our optimization method among the top 10 largest complexes and indicated in Table 1.

The exponential distribution is different from the earlier notion of the power-law degree distribution of PIN [4,5,13,21], but the power-law distribution for PIN has been contested by number of researches as well [33,34,35]. Statistically, the *diversity* of degrees and strengths in the complex-complex and protein-protein networks is less than that of PIN. Therefore, we conclude that the existence and strength of interactions with other proteins for each protein in terms of participation in protein complexes are less diverse than those for the pairwise interactions. These exponentially-decaying distributions are in contrast to the results from earlier works on biological networks [13], but recent studies suggest the different frameworks of assessing the protein interactions. It turns out that these homogeneous distributions correspond to the stable complex formation. For example, in a recent "benchmark" paper by Yu *et al*. revealed the intrinsically different natures of detected interactions from the Y2H for detecting more transient and condition-specific interactions and mass spectrometry (MS) methods for detecting stable protein complexes [34], which can support our findings of the exponential distribution of degrees, in contrast to the power-law distribution for PIN from the Y2H method. To be more specific, the detection of complexes with TAP-MS method used in our data corresponds to the MS technique, which is basically to detect "prey" proteins with the "bait" proteins so the collection of prey proteins is considered as an entity of interactome, in contrast to Y2H method using 1:1 pairwise screening. The differences between the "spoke model" (only the direct bait-prey relation is considered as interactions) and "matrix model" (all the protein pairs in bait and prey sets are interconnected) described in Ref.

[35] are also a good mechanism causing the different kind of distributions. Interestingly, there are lots of proteins having more than 100 neighbors in the projection network as shown in Fig. 3(b), while the recent PIN analysis shows the maximum degree less than 100 in spite of their power-law distribution [34]. Here we do not claim that our finding is in conflict with the previous studies on the binary PPI, but that this projection network based on TAP-MS complex data clearly reflects fundamentally *different* type of interactions in the yeast proteome, as well-documented in Ref. [34].

The exponential or at least "bounded" degree distribution of complexes in the bipartite network can be thought to originate from, for example, the "crowding effect" in a cell. Cytoplasm of a cell is occupied by various macromolecules such as enzymes, whose concentration cannot be further increased without affecting protein-protein association or transport dynamics [36]. Specifically, there should be a limit in the available cytoplasmic volume a complex can take, and thus the number of component proteins it can bear within the limited physical space. Also, it is empirically observed that the number of complexes a protein can participate in (= the protein's degree in the bipartite network) is limited as well. For the exponential degree distributions for both parties in the bipartite network, the projected networks' degree distributions actually turn out to be exponential-like unless there are nontrivial correlations, because each party's degree distribution mainly determines the projected network's degree distribution as shown in Ref. [37]. We have derived more rigorous proof of this fact by adopting a similar technique to Ref. [38], where the generating function for the degree distribution is used to derive the degree-degree correlation for projected networks. It is straightforward to calculate the generating function of the projected protein network's degree distribution given by

$$f(x) = \frac{e^{2/N}(1-e^{-1/N})^2(1-e^{-1/M})^2(1-xe^{-1/N})^4(e^{1/N}-1)}{(e^{1/N}-x)^3[1-2xe^{-1/N}+x^2e^{-2/N}-e^{-1/M}+2e^{-(N+M)/(MN)}-e^{-(N+2M)/(MN)}]^2},$$

where the degree distributions of proteins and complexes in a bipartite network are given by $p_{\text{protein}}(k) \sim \exp(-k/M)$ and $p_{\text{complex}}(k) \sim \exp(-k/N)$ and there is no degree-degree correlation ($e_{jk} = p_j p_k$, i.e., the joint degree distribution of the vertex pair $j$ and $k$ is given by the product of each vertex's degree distribution), which we have checked that as a reasonable assumption from our dataset, by the way. The resultant degree distribution of proteins in the projected network is calculated with the formula $p(k) = \frac{1}{k!}\left(\frac{d^k}{dx^k}f(x)\right)_{x=0}$ and we numerically check that the distribution is indeed very close to exponential one for $k \gg 1$, as shown in Fig. S2 (in Additional Files).

One may argue that it is hard to tell something about the distribution with the limited number of datasets, but in any case, we can safely exclude the power-law distribution from fitted curves from Figs. 2-3 and S1. Also, we emphasize that our dataset from the TAP-MS method, modified to successfully overcome the membrane protein under-representation problem [16], is fairly genome-wide. Even if there might be missing proteins or complexes as a result of experimental limitation of detecting heterodimer complexes, the degree distribution would not be severely affected by sampling. In a previous work including PIN data [39], we showed that the functional form of degree distribution is conserved under sampling up to quite a low sampling fraction. In this respect, we conclude that the different type of distribution in this case reflects the actually different characteristics of the binary interaction and the interaction as forming complexes.

To check this property of network statistics in case of other organisms, we use the list of protein complexes of *Escherichia coli* (*E. coli*) data [40] and Human Protein Reference Database (HPRD) [41]. First, since the *E. coli* dataset defined in Ref. [40] does not provide any overlapped component protein between complexes, we observe only the degree distribution of complexes, meaning the number of component proteins each complex has. It is difficult to reach a definite conclusion of the distribution's functional form with a very small range of degree values, but the maximum degree is less than 40 and the argument of crowding effect seems to hold here as well (Fig. S3 in Additional Files). In HPRD datasets, there are both the list of protein complexes and binary PPI, which allows the comparative analysis. As a result, the degree distributions of complexes and proteins in the bipartite network are less broad than the ones in the binary PPI (Fig. S4 in Additional Files), which indicates a similar result to our *S. cerevisiae* dataset. The degree distributions of the bipartite network of HPRD themselves do not seem to follow the exponential distribution, though, and we suspect that this difference may stem from the fact that HPRD dataset is basically manual literature mining of small-scale experiments, not the result from a systematic high-throughput experiment. On the other hand, if the degree distribution from HPRD reflects biological reality rather than such sampling biases, we suppose that the relatively large size of mammalian cells may "relax" the aforementioned macromolecular crowding effect, thereby enriching the tail part of the degree distribution, compared to the case of yeast cells.

**Distinction between Core and Attachment Proteins**

The list of component proteins for each protein complex is divided into two groups, *cores* and *attachments* in Ref. [16]. The "standard" complex-protein dataset is selected in terms of best accuracy and coverage in a set of manually-curated complexes, but there can be other possible variations of complex-protein sets if the threshold of accuracy and coverage for clustering becomes more flexible. These variations in reasonable ranges are denoted as "isoforms," and whether a protein is core or attachment is determined by the participation ratio in the isoforms. In other words, proteins present at a certain complex in a significant portion of the isoforms are classified as core proteins of that complex, and denoted as attachment proteins of that complex otherwise. Note that the entire set of proteins is not exactly partitioned into two groups, because a core protein for a complex can be an attachment protein in another complex. In fact, as mentioned in the 'Methods' section, 790 proteins among the total number of 1,491 proteins act as both cores and attachments. Core proteins seem to bind more rigidly, and form the basic unit of a given complex.

If only the core proteins are considered to construct the bipartite network for complexes and proteins, the average degree of core proteins in the bipartite network is 1.28, and most (about 80%) proteins participate in only one complex, while the largest degree is only 5. On the other hand, when considering only attachment proteins, the average degree of attachment proteins in the bipartite network is 4.83 and the largest degree is 24. The average degree is 4.42 [between 1.28 (only core) and 4.83 (only attachment)] in case of all proteins, and from this comparative analysis we can observe that core proteins are much less promiscuous than the attachment ones. The degree distributions of proteins for both cases are closer to the exponential distribution, as shown in Fig. 4. This quantitatively different participation pattern suggests that the core proteins in a protein complex act as its "skeleton" and the attachment proteins play the role of "seasoning," because attachment proteins are likely to participate in the formation of a relatively large number of complexes. One-mode projections, when only core proteins or only attachment proteins are considered, are shown to have exponential degree and strength distributions for both complex-mode and protein-mode projections, similar to the case without the distinction between those two kinds of proteins.

Different roles of core and attachment proteins are also reflected in the clustering structure of complex-mode projection network. If only the core proteins are considered, the complex-mode projection network is composed of 42 separate connected components, which is a highly fragmented configuration. Thus, core proteins are important as a stable component for individual complexes, rather than a mediator which "binds" complexes together. On the other hand, if only the attachment proteins are used to bond complexes, one single giant component emerges, except for a single tiny component with only two complexes. This small component is also merged into the single giant component if we consider both core and attachment proteins. As a representative example, we present the exosome 3'-5' exoribonuclease complex (complex 326 in Table S1) which contains all the core components of SKI complex (complex 364 in Table S1) as attachments, supporting previous works that the association is necessary for cytoplasmic messenger RNA 3'-to-5' decay [16,42]. Therefore, the attachment proteins (core proteins of SKI complex in this case) indeed specify a particular function for the exosome complex.

**Inference of Abundance and Functions: Statistical Validation**

There are core and attachment proteins, as stated and analyzed in the previous subsection. To make our predictions as *conservative* as possible, we adopt the following scheme about selecting core and attachment component proteins. For the abundance estimation, we consider only the core proteins as components of complexes, based on the fact that the core proteins play the role of the skeletons as previously mentioned, *i.e.*, unlike the attachment proteins shared by relatively large number of complexes which can overestimate the number of complexes. In other words, we expect that utilizing only the core proteins leads us to a more accurate estimate of the abundance. In contrast, both the core and attachment proteins are used in function assignment of complexes. In this way, it allows more alternative ways of function assignments, and if a function is assigned in spite of all those possibilities, we can consider it to be a genuine function, with more certainty. However, in case of assignment of previously unknown functions to proteins after the function assignments for complexes, only core proteins are considered instead, reflecting the fact that the proteins' main biological functions are achieved by participating as core components. Finally, we remark that the $\{S_{ij}\}$ matrix for abundance estimation is approximated by $\{U_{ij}\}$ matrix, due to the limitation of available information. However, note that more and more accurate datasets will become available in the future, and applying our method for those new datasets would be straightforward.

With the LP and MIP method, we have numerically solved the optimization problems to obtain the abundance [minimizing Eq. (2)] and functions [minimizing Eq. (4)] of complexes and their component proteins. As the input datasets for individual proteins, we use condition-dependent abundance (arbitrary units) of proteins [27] and the MIPS functional database for individual proteins [28]. In total, 325 different MIPS FunCat functions are assigned to our individual protein set. Since the MIPS FunCat classifications are hierarchically organized [28], we can divide the set of functions into more coarse-gained "functional categories" from their higher level hierarchy to observe more general trends, as demonstrated in the next paragraph. Before we apply the method for the main result, we have divided the known input datasets into several training and test sets and verified its effectiveness for various settings. The full results for our function assignments for both complexes and proteins are available (Tables S1 and S2 in Additional Files). Note that we also present an alternative optimization method in Additional Files, denoted as 'Gavin_2nd' in Tables S1 and S2, minimizing the number of proteins for each newly assigned function, instead of the number of complexes described in Eq. (4) (denoted as 'Gavin_1st'). The MIP method does not produce a unique solution in general, and there can be multiple numbers of solutions. Among those solutions, we select high confidence (HC) outcomes among the original outcomes (denoted as "raw outcomes") from our method. In other words, HC outcomes contain only the functions which are assigned to complexes and proteins, in all the multiple solutions.

To check the accuracy of our method, we first divide the set of proteins with known values of abundance into training and test sets. Assuming that the values of abundance for the proteins in the test set are unknown, we estimate the abundance for those proteins in case of rich media (YEPD) [27] and compare them with real values. We define the relative deviation as

$$\alpha = |[\text{estimated abundance}] - [\text{real abundance}]| / [\text{real abundance}]. \qquad (6)$$

Then, we plot the ranked $\alpha$ values, compared to the ones for random pairing of estimated abundance and real abundance in Fig. S5 (in Additional Files). The real deviation values are always significantly smaller than those of random counterparts outside the error range, demonstrating the statistical significance of our results. We have also observed the similar result for minimal media (SD) [27] as well.

For the function assignment, we also divide the set of proteins with already-known functions into training and test sets. Assuming all the protein functions in the test set are unknown, we assign those functions based on our function assignment method; the results are shown in Table 2. We measure two complementary quantities to show both sensitivity and specificity, for both the raw outcomes and HC outcomes. From the results, we conclude that our method works reasonably well, especially for HC outcomes, considering the noise level of this kind of high-throughput dataset. One remarkable thing is that our method works quite well for a very small fraction of training set such as 20% for both abundance estimation and function assignment, and this fact implies that highly interconnected relationship among proteins via complex membership greatly helps us to correctly assign functions. To compare our results with a recent function assignment scheme, we adopt the hypergeometric test used in Ref. [23] ('CYC2008' dataset) and applied the method with the same threshold (E-value = 0.05) to our dataset. Note that the result obviously depends on the threshold value, and also the original CYC2008 method does not include the "re-substitution" process of copying functions from complexes to their components. Therefore, we have to combine our re-substitution process after the CYC2008 procedure for function assignment for complexes. The compared results (shown in Table S3 in Additional Files) clearly show that the performance of our method is better than or at least comparable to that of CYC2008 in many cases, and especially the number of predicted functions itself is much smaller for CYC2008 method. To be more specific, 'fraction of proteins for our high-confidence (HC) result' shown in the third column is always better than the CYC2008 result in the fourth column, and even our raw result in the second column is better than CYC2008 in case of large values of training set fraction $p$ or at least comparable to each other for small $p$. The only case worse than CYC2008 is the 'fraction of functions for our raw result' in the fifth column for small $p$, but this can also be overcome by using HC dataset in the sixth column. Therefore, we conclude that our global optimization method is much better at guessing the unknown protein functions, in terms of sensitivity or recall rate, let alone the problem of dependence on an arbitrary parameter, which is the wholesale E-value threshold for the CYC2008 method.

**Inference of Abundance and Functions: Condition-Dependent Abundance with Regard to Functional Categories**

First, we show the average change of complexes' abundance depending on the condition (rich or minimal media), classified as each MIPS functional category in Fig. 5. Note that a complex can be assigned with multiple functions, so there are overlaps among different functional categories. We believe that this cross-analysis of abundance and function assignment strongly supports the reliability of our method, in addition to the statistical test presented in the previous subsection.

From Fig. 5, one can check that each functional category shows a unique pattern of the condition-dependent abundance changes. The most notable case is the increased abundance of complexes doing metabolism in the minimal media compared to the rich media, indicating that many proteins induced in minimal medium are involved in the production of small molecules that cannot be taken up from the surrounding environment. Another notable category of increased abundance in the minimal media is the cellular communication. For instance, the protein kinase complex 312 in Table S1 is significantly more abundant in the minimal media, and this result can be interpreted from the nutrient sensing role of component protein Tpk2 involved in the cell growth via the Ras-cAMP signaling pathway, at facing the nutrient starvation [28]. Also, the abundance of complex 77 in Table S1 is increased in the minimal media and one possible explanation is its component protein transketolase's (Tkl1's) involvement in the pentose phosphate pathway of processing glucose 6-phosphate to produce ribose-5-phosphate, which can be demanded more in the minimal media [28].

In contrast, many complexes with functions of cell cycle and protein synthesis tend to show decreased abundance in minimal media, indicating that many proteins induced in rich media are involved in cell growth and division. A fraction of the complexes with each functional category, for three different ranges of abundance ratio changes, shown in Fig. 6, also shows a similar trend. Our findings are consistent with the argument in Ref. [27], which discusses such effects on the individual protein level. Note that, however, the previous work in Ref. [27] is about individual proteins, while our results are about complexes, by which the real biological functions are expressed. It indicates that the similar arguments are applied to abundance and functions of complexes.

**Inference of Abundance and Functions: Other Examples**
From our results about the functional annotations for complexes, we find a number of intriguing examples. For instance, both the complex 276, whose core proteins are Ste11 and Ste50, and the complex 312, whose core proteins are Bcy1 and Tpk2, are assigned with the cytoskeleton formation and signal transduction. The relationship between the cytoskeleton proteins and the cell signal transduction is discussed in recent literatures [43,44], arguing that the filaments of the cytoskeleton provide guiding tracks so that transport becomes more focused. The core component proteins of those complexes we have found, Ste11, Ste50, Bcy11, and Bcy1, are also shown in Ref. [43]. Another example is related to the shifting process from fermentation to respiration in yeast [45], depending on the temporal order of environmental stimuli, such as osmotic pressure and heat shock. As shown in Table S1, there are complexes co-assigned with those functions related to the process such as complex 14, 121, 212 (carbohydrate metabolism, and osmotic and salt stress response), and complex 79 (aerobic respiration and heat shock response).

Our optimization scheme does not only give the systematic estimate of complexes' abundance and function assignment, but also conjecture previously unknown new functions to individual proteins, as stated in the 'Methods' section. Table 3 shows biologically interesting examples from our results with manually curated references for validation. For instance, the metabolic enzyme Eno2 was not assigned with any functions related to the cell division, but assigned with the cell division functions as a result from our optimization technique. An important point here is that, as we will demonstrate with Eno2 in the next paragraph, we know exactly where the functions come from (at least in our optimization scheme), *i.e.*, by actually tracing the complex to which the function is assigned and looking for another component protein participating in the complex, as the "source" of the function.

In case of the protein Eno2, it participates in the complex 289 (the same complex index is used as in Ref. [16]) as a core component, and the complex is involved in the cell division process (10.03.03 in the MIPS classification scheme). The function assigned to that complex originates from another core component protein Chs1 in the MIPS datasets. The protein Eno2's involvement in the cell division is also supported by a recent experiment [46], showing that the overexpression of the protein significantly delays the progression of yeast cell cycle. Of course, the experimental results need to be carefully interpreted, since G1 arrest can also be made by perturbing proteins not directly involved in the cell division process, as exemplified by Sko1 activating the pheromone response pathway [46]. This type of evidence, therefore, may not be a *sufficient* condition for the direct involvement in the cell division process, but can be a *necessary* condition which hints the candidate targets for further investigation in a more elaborate way. With such procedure, we find quite a number of examples indeed, as shown in Table 3, and suggest small-scale experiments to confirm the functional annotations of these proteins.

## Discussion

It is indispensable for the future of proteomics to understand the composition structure and interactions of protein complexes. However, relatively fewer works about the protein complexes have been done in the genome-wide level, except for a few cases [14-18], compared to works about pairwise PPI. In this work, we have investigated and shown the applications of the interactions in the recently published protein complex database of the budding yeast [16]. We have adopted the notion of bipartite network, which is suitable for the case where there are two kinds of elements and one of them acts as the set of components of the other. By constructing the bipartite network with protein complexes and their component proteins, we have checked the degree distributions of complexes and proteins.

As results of the analysis, the degree distribution of the bipartite network asymptotically follows the exponential distribution. The same is true for the degree and strength distributions of the one-mode projection networks as well. Interestingly, the resultant exponential degree distributions are different from the heavy-tailed distribution observed in most pairwise PPI networks. Many complexes and proteins with largest degrees and strengths are affiliated to the ribosome, allowing for the stable formation of large ribosomal complexes. Furthermore, we have shown that the core and attachment proteins in the datasets show different participation patterns in the formation of complexes, which illustrates the different roles in the complexes.

The most valuable contribution of our work is presenting the new optimization method without external tuning parameters to determine the abundance and function of protein complexes, along with the previously unknown properties of individual proteins, based on the information of their global organization. The effectiveness of our optimization method is proved by various biologically relevant examples compared with experimental results. Since we provide the full results from our results as Additional Files, we hope our result can encourage other researchers to utilize it as their new dataset for deeper understanding of proteomics.

## Conclusions

Thanks to the rich proteomics datasets of *S. cerevisiae*, many genome-wide researches have been achieved. Analyses based on the protein complexes can open a door to a new understanding of how proteins interact and work in organisms. Important future works include more investigation about our bipartite (and one-mode projections) network topology, along with more systematic comparison with the conventional PIN to see the detailed origin of the different form of the degree distribution. Ultimately, it is essential to perform the small-scale experiments to validate the results claimed by our new optimization method. We hope this work can make a substantial contribution to this direction of research.

## Competing interests

The authors declare that they have no competing interests.

## Authors' contributions

SHL, PJK, and HJ conceived and designed the study. SHL collected the data. SHL and PJK performed the numerical simulation and statistical analysis, and analyzed the results. SHL, PJK, and HJ wrote, read, and approved the final manuscript.

## Acknowledgements

This work is supported by the Swedish Research Foundation (SHL) and the Korean Systems Biology Research Project (20100002164) of the Ministry of Education, Science and Technology (MEST) through the National Research Foundation of Korea (HJ). The authors greatly acknowledge Sangjo Han, Yong-Su Jin, Carl T. Bergstrom, and Petter Holme for fruitful discussions and thoughtful comments. We also thank Sejun Lee, Juhyun Jung, and Doheon Lee for providing information about the previous works, and Jimin Hwang and Seung Ki Baek for their linguistic advice.

# Figures

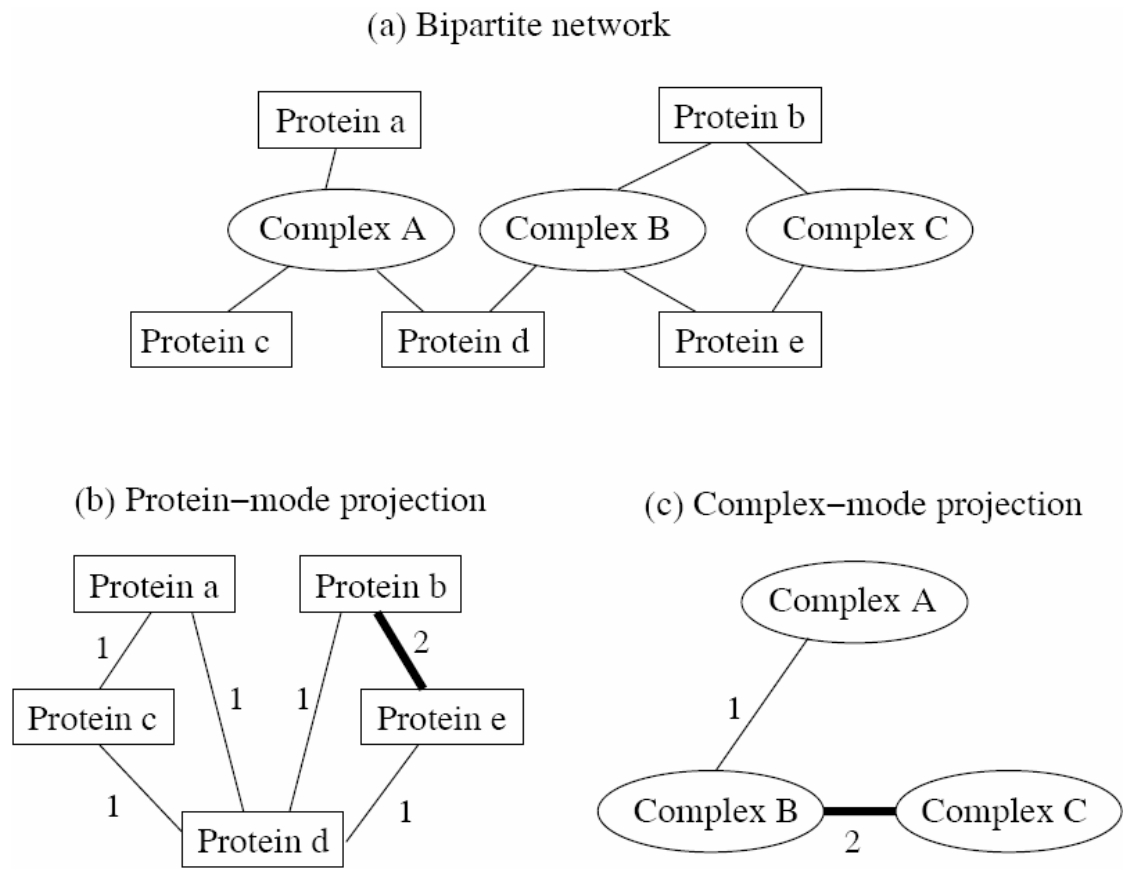

**Figure 1 - An example of bipartite network representation of protein complexes and component proteins**

(a) Complex A is composed of protein a, b, and c, complex B of protein b, d, and e, and complex C of protein b and e. (b) In protein-mode projection, two proteins are linked if they share a complex and the link's weight is assigned as the number of complexes they share. (c) Similarly, in complex-mode projection, two complexes are linked if they share a component protein and the weight is given by the number of proteins they share.

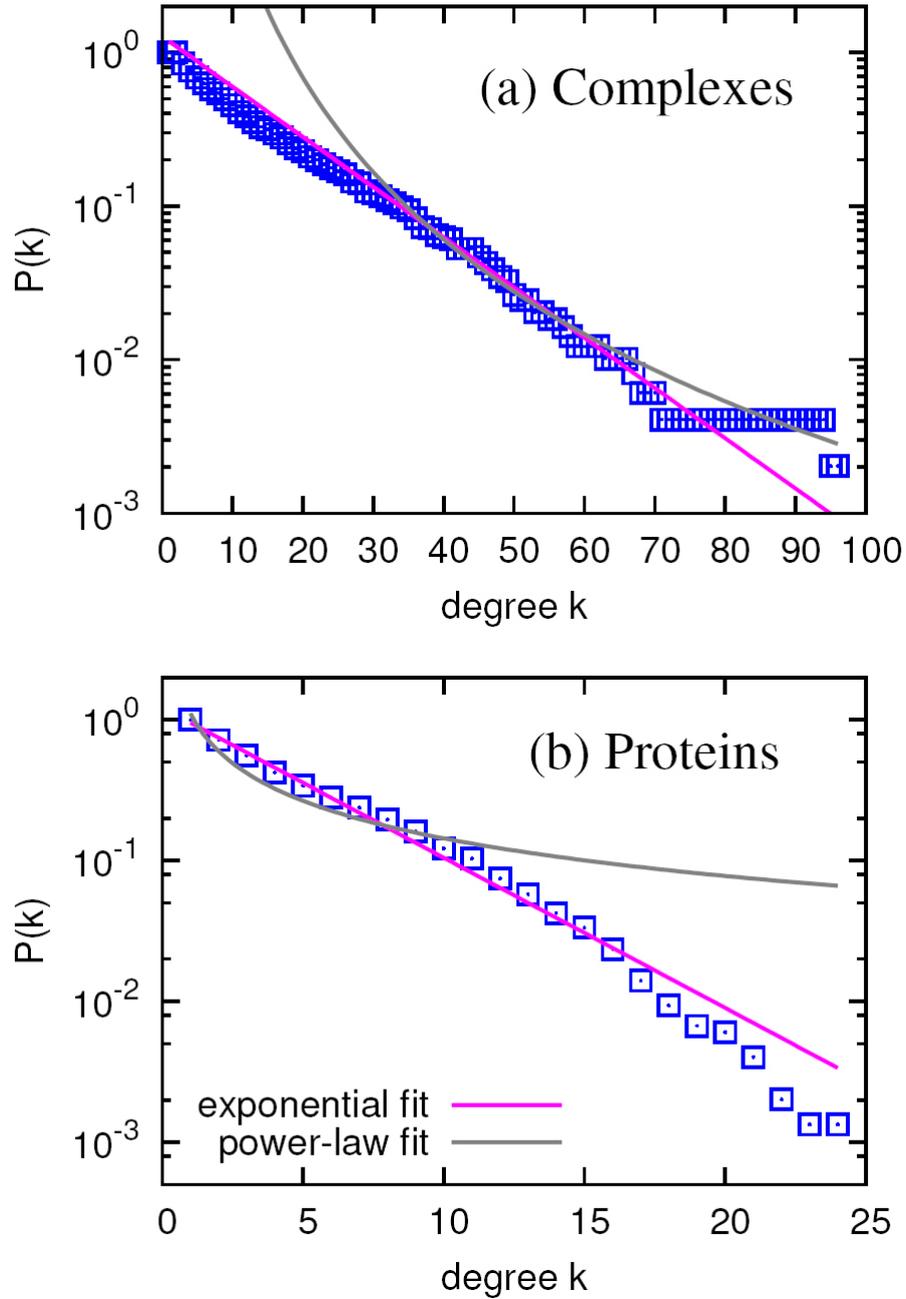

**Figure 2 - Degree distributions of the bipartite network, for (a) complexes and (b) proteins**

Here, the degree in (a) corresponds to the number of component proteins for each complex, and the degree in (b) corresponds to the number of complexes in which a protein participates as a component. The blue squares correspond to the cumulative degree distribution $P(k) = \sum_{k' \geq k} p(k')$, and the pink lines and gray curves are the best exponential and power-law fittings, respectively.

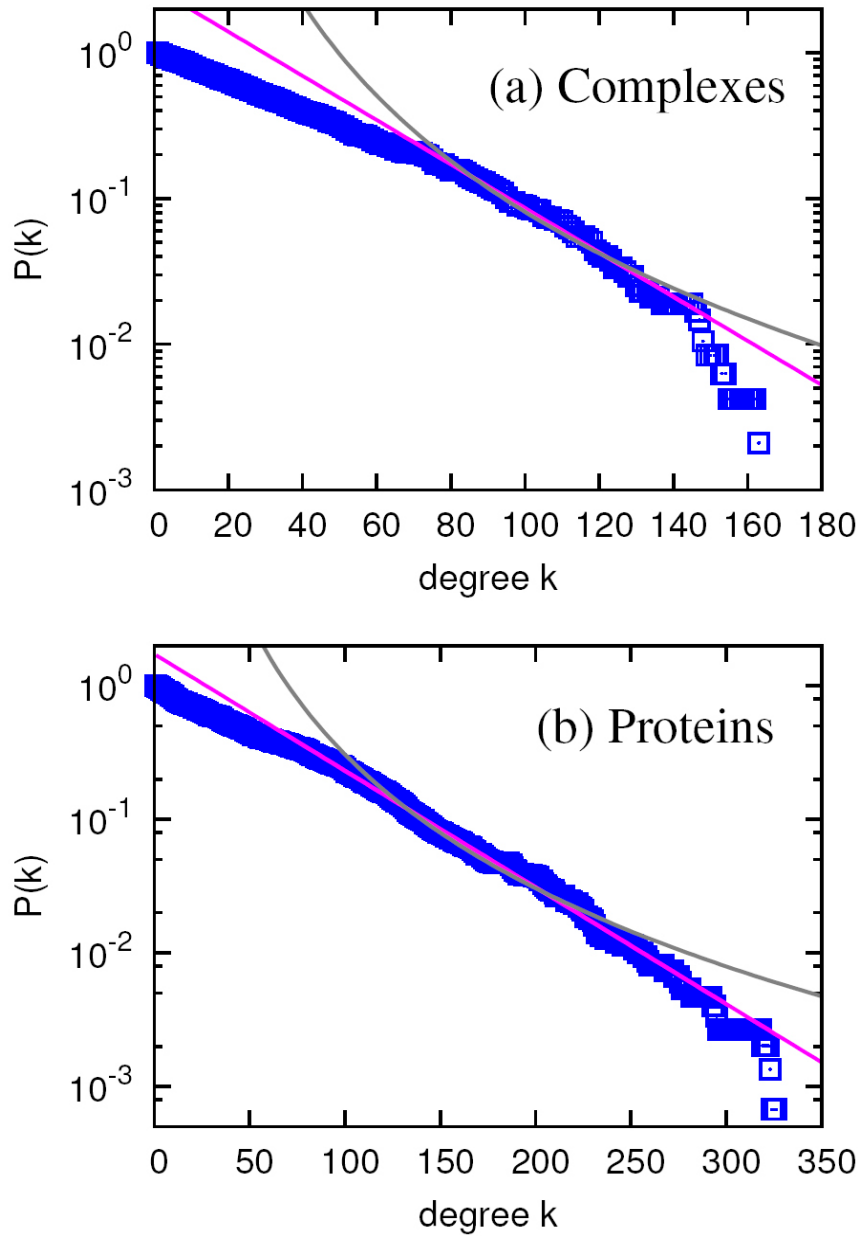

**Figure 3 - Degree distributions of (a) complex-mode projection (complex-complex network) and (b) proteins-mode projection (protein-protein network)**
Here, the degree in (a) corresponds to the number of complexes which share at least one component protein with each complex, and the degree (b) corresponds to the number of proteins which participate in at least one complex with each protein. The blue squares correspond to the cumulative degree distribution $P(k) = \sum_{k' \geq k} p(k')$. The pink lines and gray curves are the best exponential and power-law fittings, respectively.

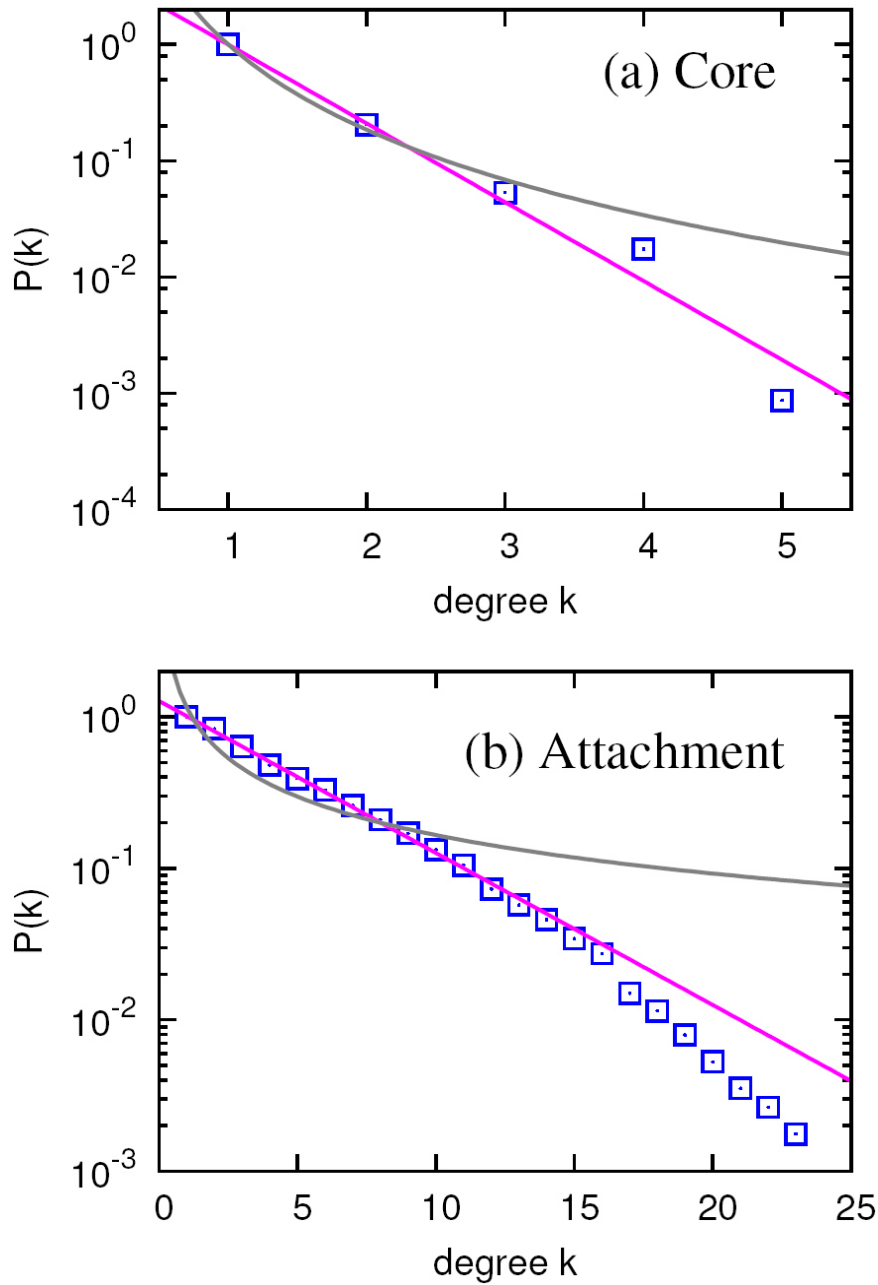

**Figure 4 - Degree distributions of proteins for the bipartite network if (a) only core proteins are considered and (b) only attachment proteins are considered**
Here, the degree in (a) corresponds to the number of complexes in which a protein participates as a *core* component, and the degree in (b) corresponds to the number of complexes in which a protein participates as an *attachment* component. The blue squares correspond to the cumulative degree distribution $P(k) = \sum_{k' \geq k} p(k')$. The pink lines and gray curves are the best exponential and power-law fittings, respectively.

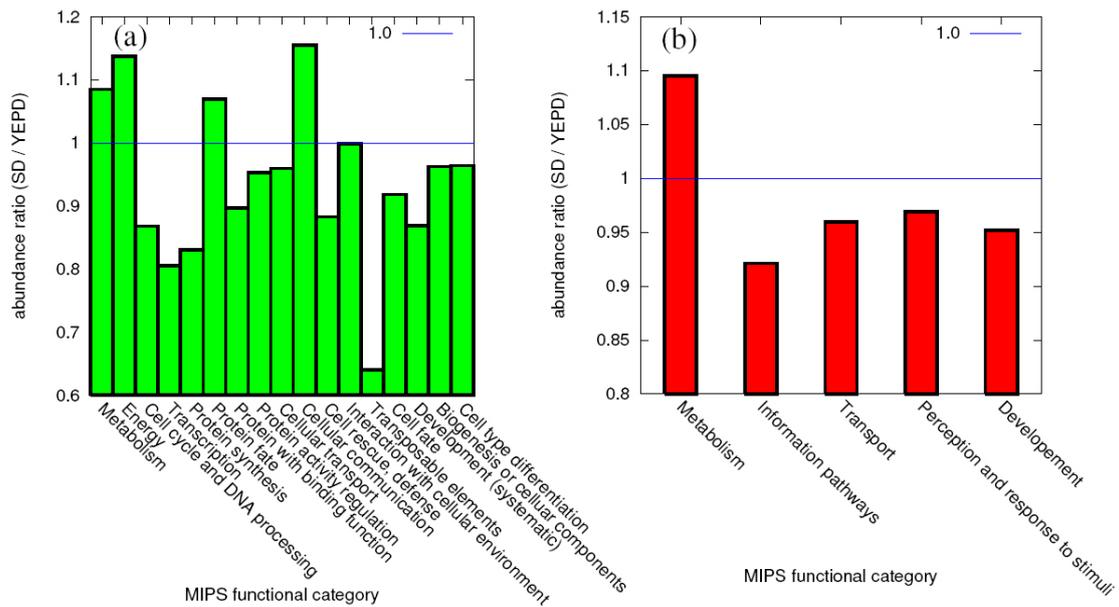

**Figure 5 - Abundance ratio of complexes for each MIPS functional category, classified as (a) main functional categories and (b) coarser classification.**
The abundance ratio in y-axes is defined as the ratio of the abundance in minimal (SD) media environment to the abundance in rich (YEPD) media. High confidence (HC) outcomes are used.

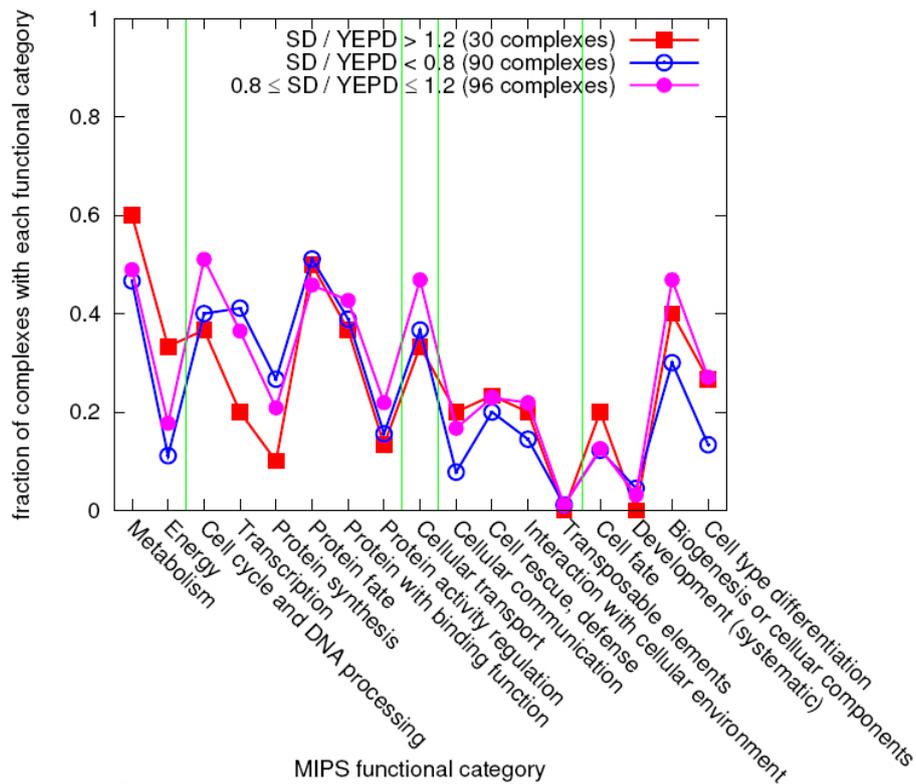

**Figure 6 - Fraction of complexes with each functional category**

The complexes are divided into three sets according to their abundance changes (red for > 120%, blue for < 80%, and pink otherwise) depending on the media. For each functional category, the fraction of complexes including at least one function corresponding to the category is shown, for each abundance change classes. Light green vertical lines divide the coarser functional categories shown in Fig. 5(b). Raw outcomes are used.

# Tables

**Table 1 - List of complexes and proteins with ten largest degree/strength in bipartite/projection networks**

| Complex in bipartite network (degree) | Protein in bipartite network (degree) | Complex in projection network (degree) | Complex in projection network (strength) | Protein in projection network (degree) | Protein in projection network (strength) |
|---|---|---|---|---|---|
| 27• (94) | Rps22a (24) | 27• (163) | 56• (1047) | Rpl33a (326) | Utp20 (860) |
| 56• (94) | Rpl36b (24) | 56• (162) | 27• (994) | Rpl30 (323) | Rpl33a (815) |
| 37 (65) | Rpl11b (22) | 10••• (154) | 55• (722) | Rpl16b (322) | Nop1 (800) |
| 55• (65) | Rpl26b (21) | 32** (152) | 37 (680) | Utp20 (318) | Rpl26b (790) |
| 1* (61) | Nop1 (21) | 58 (148) | 18•• (669) | Rps14a (295) | Rps22a (721) |
| 18•• (58) | Rps14a (20) | 37 (147) | 23 (656) | Rps22a (294) | Rpl11b (711) |
| 23 (57) | Rp130 (20) | 45†† (147) | 331 (640) | Nop1 (292) | Mpp10 (698) |
| 39 (56) | Rp12b (20) | 18•• (146) | 371 (581) | Sro9 (281) | Rpl36b (678) |
| 80† (55) | Utp20 (20) | 331 (145) | 39 (577) | Rrp12 (276) | Rpp2b (675) |
| 10••• (52) | Rpl33a (16) | 41 (136) | 211 (530) | Rpl6b (275) | Rps14a (673) |

The annotated complexes in Ref. [16] are following: *U1 snRNP complex, †Prp19-associate complex, **ribosome-associate complex (RAC), ††Translation initiation factor eIF3 complex, •complexes assigned as 'ribosomal proteins' function, ••tRNA modification function, and •••fatty acid metabolism and aromate anabolism predicted by our optimization method (Table S1).

**Table 2 - Statistical validation of the function assignment**

| $p$ | Fraction of functions* (raw) | Fraction of functions* (HC) | Fraction of proteins† (raw) | Fraction of proteins† (HC) |
|---|---|---|---|---|
| 0.20 | 33.5% (641/1912) | 50.4% (502/996) | 61.6% (278/451) | 73.5% (219/298) |
| 0.40 | 25.8% (645/2500) | 40.0% (532/1330) | 64.3% (270/420) | 71.4% (227/318) |
| 0.50 | 26.7% (608/2275) | 41.6% (480/1153) | 62.4% (232/372) | 72.1% (191/265) |
| 0.60 | 28.2% (583/2068) | 44.0% (497/1129) | 68.3% (213/312) | 74.2% (187/252) |
| 0.80 | 19.3% (274/1418) | 32.7% (231/706) | 66.3% (110/166) | 75.8% (97/128) |
| 0.90 | 23.3% (147/630) | 37.5% (135/360) | 71.3% (57/80) | 78.8% (52/66) |
| 0.95 | 18.2% (78/429) | 32.8% (67/204) | 59.2% (29/49) | 65.8% (25/38) |
| 0.98 | 23.1% (30/130) | 39.2% (29/74) | 56.3% (9/16) | 75.0% (9/12) |
| 0.99 | 10.6% (14/132) | 18.6% (11/59) | 63.6% (7/11) | 60.0% (6/10) |

Here $p$ is the fraction of training set. *Among all the newly assigned functions to the entire test set proteins, fraction of functions which are also on the original MIPS functional datasets. †Fraction of test proteins with at least one function assigned, which has at least one assigned function on the original MIPS functional datasets. Raw outcomes correspond to the original outcomes from our method, and high confidence (HC) outcomes are the selected subset of them whose function assignments are invariant even if the multiple solutions are considered.

**Table 3 - Some notable examples of proteins with newly assigned biological functions with our optimization method**

| Protein | Known function | Predicted function | Reference |
|---------|----------------|--------------------|-----------|
| Eno2 | Metabolism | Cell division | [46] |
| Pgk1 | Metabolism | Cell division | [47] |
| Reg1 | Metabolism/regulation/cell growth/morphogenesis | Cell division | [48] |
| Act1 | Cell cycle/growth | Mitochondrial biogenesis | [49] |
| Gsg1 | ER to Golgi transport | Cell wall biogenesis | [50] |
| Ydl203c | Metabolism | Cell wall biogenesis | [46] |
| Ymr237w | Metabolism | Cell wall biogenesis | [46] |
| Pep3 | Protein targeting, sorting and translocations, vacuolar/lysosomal transport | Cell wall biogenesis | [51] |
| Ubp15 | Cytoplasmic and nuclear protein degradation | Cell wall biogenesis | [52] |
| Atg17 | Cytoskeletal biogenesis | Peroxisome biogenesis | [53] |
| Gpi8 | Lipid, protein modification | Endoplasmic reticulum biogenesis | [54] |
| Yol070c | Unknown | Bud/growth tip, cell cycle checkpoint | [29] |
| Hda2 | Transcriptional control | Metabolic enzyme | [55] |
| Iwr1 | Meiosis | Transcriptional control | [29] |

Reference [29] refers to the case that the predicted functions are not given by the MIPS data [28] so that the functions are not assigned initially, but found by our method and on the Gene Ontology (GO) database [29]. In this case, we consider GO database as the validation source, similar to other references [46-55].

## Additional files

**Additional file 1 – List of all the protein complexes with MIPS functions assigned by our method (Table S1)**

List of all the protein complexes with MIPS functions assigned by our method, where the core and attachment components are taken from Ref. [16] (and the indices are the same as Ref. [16]). We classify each function into the following three categories. (1) Gavin_1st (light yellow): minimizing the number of complexes for each newly assigned function, described as Eq. (4) in the main text. (2) Gavin_2nd (light green): minimizing the number of proteins for each newly assigned function, instead of that of complexes (3) The functions assigned by both (1) and (2) (bright yellow). Note that we only select high confidence (HC) outcomes among the raw outcomes, whose reliability of function assignment considering the multiple solutions is large.

**Additional file 2 – List of all the proteins with MIPS functions newly assigned by our method (Table S2)**

List of all the proteins with MIPS functions newly assigned by our method (bright yellow), along with the ones from the MIPS database (sky blue) which is used as "input function." HC outcomes are selected as in Table S1. Sometimes the functions already annotated in MIPS and the ones assigned by our method are quite similar, and very different in some cases. Therefore, we suggest the latter case be worth investigating further, as we did in the main text.

**Additional file 3 – Statistical validation of the function assignment, in comparison with the hypergeometric test in Ref. [23] (CYC2008) (Table S3)**

Here $p$ is the fraction of training set. $^{\dagger}$Fraction of test proteins with at least one function assigned, which has at least one assigned function on the original MIPS functional datasets. $^{*}$Among all the newly assigned functions to the entire test set proteins, fraction of functions which are also on the original MIPS functional datasets. Raw outcomes correspond to the original outcomes from our method, and high confidence (HC) outcomes are the selected subset of them whose function assignments are invariant even if the multiple solutions are considered.

### Additional file 4 - Strength distributions of (a) complex-mode projection and (b) proteins-mode projection (Figure S1)

Here, the strength in (a) corresponds to the sum of number of proteins shared with the neighboring complexes for each complex, and the strength (b) corresponds to the sum of number of complexes shared with the neighboring proteins for each protein. The blue squares correspond to the cumulative strength distribution $P(s) = \sum_{s' \geq s} p(s')$, and the pink lines and gray curves are the best exponential and power-law fittings, respectively.

### Additional file 5 - Projected network's degree distribution from the generating function approach (Figure S2)

Assuming that both protein and complex's degree distributions follow the exponential degree distribution $p(k) \sim \exp(-k)$, the projected network's degree distribution is numerically calculated with the generating function approach mentioned in the main text. One can clearly observe that the degree distribution follows the exponential tail for $k \gg 1$.

### Additional file 6 - Degree distribution of complexes in the bipartite network for *E. coli* protein complex data in (a) semi-log scale and (b) double-log scale (Figure S3)

Here, the degree corresponds to the number of component proteins for each complex.

The blue squares correspond to the cumulative degree distribution $P(k) = \sum_{k' \geq k} p(k')$.

### Additional file 7 - Degree distribution of complexes and proteins in the bipartite network for human protein complex data, in comparison with the protein-protein interaction network (Figure S4)

Here, a complex's degree (blue square) corresponds to the number of component proteins for each complex, and a protein's degree (red circle) corresponds to the number of complexes in which a protein participates as a component. The degree distribution of proteins in the binary protein-protein interaction is shown as pink triangles, as a comparison.

**Additional file 8 - Statistical validation of abundance estimation (Figure S5)**

For each value of fraction of training set $p$, the relative deviation

$$\alpha = |[\text{estimated abundance}] - [\text{real abundance}]| / [\text{real abundance}]$$

is ranked and shown compared to its random counterparts (meaning that the identity of proteins with estimated abundance is randomly paired with the ones with real abundance values). The real deviation values are always on the left side (smaller than) of the random counterparts outside the error range, which implies its statistical significance.